\documentclass[prd,twocolumn,superscriptaddress,nofootinbib,showpacs]{revtex4}
\usepackage{graphicx,amsmath,amsfonts,amssymb,revsymb,dcolumn,epsfig,bm}

\begin{document}

\title{Circles-in-the-sky searches and observable cosmic topology in the
inflationary limit}

\author{B. Mota}\email{wronski@gmail.com}
\affiliation{Centro Brasileiro de Pesquisas F\'{\i}sicas,
Rua Dr.\ Xavier Sigaud 150 \\
22290-180 Rio de Janeiro -- RJ, Brazil}

\author{M.J. Rebou\c{c}as}\email{reboucas@cbpf.br}
\affiliation{Centro Brasileiro de Pesquisas F\'{\i}sicas,
Rua Dr.\ Xavier Sigaud 150 \\
22290-180 Rio de Janeiro -- RJ, Brazil}

\author{R. Tavakol}\email{r.tavakol@qmul.ac.uk}
\affiliation{Astronomy Unit, School of Mathematical Sciences,
Queen Mary, University of London\\ Mile End Road, London, E1 4NS, UK}

\date{\today}

\begin{abstract}
While the topology of the Universe is at present not specified by
any known fundamental theory, it may in principle be determined
through observations. In particular, a non-trivial topology will
generate pairs of matching circles of temperature fluctuations in
maps of the cosmic microwave background, the so-called
circles-in-the-sky. A general search for such pairs of circles
would be extremely costly and would therefore need to be confined
to restricted parameter ranges. To draw quantitative conclusions
from the negative results of such partial searches for the
existence of circles we need a concrete theoretical framework.
Here we provide such a framework by obtaining constraints on the
angular parameters of these circles as a function of cosmological
density parameters and the observer's position. As an
example of the application of our results, we consider the recent
search restricted to pairs of nearly back-to-back circles with
negative results. We show that assuming the Universe to be very
nearly flat, with its total matter-energy density satisfying the
bounds $0<|\Omega_{0}-1|$ $\lesssim10^{-5}$, compatible with the
predictions of typical inflationary models, this search, if
confirmed, could in principle be sufficient to exclude a
detectable non-trivial cosmic topology for most observers. We
further relate explicitly the fraction of observers for which this
result holds to the cosmological density parameters.
\end{abstract}


\maketitle

\section{Introduction}

One of the central open questions regarding our understanding of the
Universe concerns its shape (topology), and in particular whether it is finite
or infinite (see, e.g., the reviews~\cite{CosmTopReviews}).
A major difficulty in this regard is that general relativity,
being a metric theory, does not specify the topology of the Universe.%
\footnote{In this work, in line with the usage in the literature, by
topology of the Universe we mean the topology of its three dimensional
spatial sections $M$.}
Despite our present-day inability to predict the topology of the
Universe from a fundamental theory, it may in principle be
determined through observations. This requires high resolution
observations, which have increasingly become available in recent
years. Most notably, the on-going accumulation of data by the
Wilkinson Microwave Anisotropy Probe (WMAP)~\cite{wmap} and other
Cosmic Microwave Background (CMB) surveys have, on the one hand,
provided strong support for the inflationary
scenario~\cite{Linde2}, and  the very near flatness of the
Universe, and on the other hand made it feasible to perform
systematic searches for possible evidence of
a non-trivial topology of the Universe~\cite{Cornish-etal03,CitSRedux,%
Roukema-etal2004,Aurich-etal2006} (see also the related
Refs.~\cite{RelatedCirc}).

In the context of general relativity, the observable Universe
seems to be well described by a $4$-manifold
$\mathcal{M} =\mathbb{R}\times M$ with locally homogeneous and
isotropic spatial sections $M$ and endowed with a
Robertson-Walker metric
\begin{equation}
ds^{2}=-dt^{2}+a^{2}(t)\left[ d\chi^{2}+S_k^{2}(\chi)(d\theta
^{2}+\sin^{2}\theta d\phi^{2}) \right]  \;, \label{RWmetric}
\end{equation}
where $a(t)$ is the scale factor and $S_k(\chi)$ takes the forms $\chi\,,$
$\sin\chi\,,$ or $\sinh\chi$ depending upon whether the geometry of the
spatial sections is euclidian, spherical or hyperbolic with curvature
parameters $k=0,\pm 1$. These geometries in turn are determined by
finding out whether the total energy-matter density of the
Universe, $\Omega_0$, is equal to, greater than or smaller than 1.
Often the homogeneous and isotropic spatial sections $M$ are
assumed to be the simply connected $3$-manifolds: euclidian $\mathbb{R}^{3}$,
spherical $\mathbb{S}^{3}$, or hyperbolic $\mathbb{H}^{3}$.
However, given the metrical (local) nature of general relativity,
they can also be multiply connected $3$-manifolds (which we assume
to be compact and orientable) $M=\widetilde{M}/\Gamma$, where the
covering space $\widetilde{M}$ is respectively
$\mathbb{R}^{3}$, $\mathbb{S}^{3}$ or $\mathbb{H}^{3}$ depending
on $k$, and $\Gamma$ is a discrete and fixed point-free group of
isometries of $\widetilde{M}$ called the holonomy group.
The local geometry of the spatial sections $M$ thus constrains,
but does not dictate, its topology.

The immediate observational consequence of such multiple
connectedness is that an observer could potentially detect
multiple images of radiating sources. In particular, in a universe
with a detectable non-trivial topology the  last scattering
surface (LSS) intersects its topological images in the so called
circles-in-the-sky~\cite{CSS1998}, i.e., pairs of matching circles
of equal radii, centered at different points of the LSS with the
same distribution of temperature fluctuations along both circles.
In this way, to observationally probe a non-trivial topology on
the largest available scales, one needs to scrutinize the CMB
sky-maps in order to extract such correlated circles in order to
determine the topology of the Universe. Thus, a detectable
non-trivial cosmic topology is an observable attribute, which can
be probed through the circles-in-the-sky for all locally
homogeneous and isotropic universes with no assumptions on the
cosmological density parameters.

The conditions for the detectability of cosmic topology were
studied in Refs.~\cite{TopDetec} for classes of hyperbolic and
spherical manifolds, as a function of cosmological density
parameters. These studies were extended to the case of
\emph{generic} manifolds and general qualitative results were
obtained for the \emph{inflationary limit} in
Ref.~\cite{Mota-etal} (see also Ref.~\cite{Mota-etal-b}).
Furthermore, the inverse question of whether the detection of a
non-trivial cosmic topology can be used to set constraints on
cosmological density parameters has been studied for
\emph{specific} topologies ~\cite{Previous,Related}.

Our aims here are twofold. Firstly, to extend the previous general
results by the authors~\cite{Mota-etal}, in order to give, for
\emph{generic} detectable non-trivial topologies, a concrete
relationship between the angular parameters associated with
the circles-in-the-sky and the cosmological density parameters
in the \emph{inflationary limit}.
Secondly, by using these relations, and taking into account the observer
positions, we provide a set up to interpret the negative results of recent
and future searches for circles-in-the-sky in the WMAP data. In this way,
we can concretely specify, for example, the extent to which the recent
searches for circles-in-the-sky may be used to exclude detectable
non-trivial cosmic topologies for most observers.

The structure of the paper is as follows. In Section~\ref{Sec2} we give a brief
account of the  pre-requisites necessary for the following sections,
including a brief discussion of the so called inflationary limit, which
we use to obtain bounds on the detectability of cosmic topology.
In Section~\ref{Sec3} we combine these bounds with our previous results to
obtain bounds for the parameters of detectable holonomies in the inflationary
limit.  We quantify these bounds by giving a concrete measure of the observers
for whom the detectability holds.
In Section~\ref{Sec4} we recast our bounds on detectable holonomies in
terms of bounds on the angular parameter that measures the deviation
from antipodicity in pairs of matching circles generated by detectable
holonomies. These bounds are then compared in Section~\ref{Sec5} to the
parameters adopted in the recent searches for circles-in-the-sky to determine
whether and under which conditions it is possible to rule out a detectable
non-trivial cosmic topology given the inflationary limit. Finally, in
Section~\ref{Sec6}, we conclude with a brief discussion of the significance
of our results to draw quantitative conclusions from such negative results
of searches for circles.

\section{Preliminaries}
\label{Sec2}

A natural way to study the detectability of a topology is through the lengths
of its closed geodesics. For a holonomy $\gamma\in\Gamma$ and a point $x\in
M$, the length of the closed geodesic generated by $\gamma$ is given by its
distance function $d(x,\gamma x)$ in the covering space, i.e. the distance
between $x$ and its image $\gamma x$. This readily allows the definition of
the local injectivity radius $r_{inj}(x)$ as half the length of the smallest
closed geodesic passing through the point $x$.

A necessary condition for detectability of cosmic topology is then
given by
\begin{equation}
r_{inj}(x)<\chi_{obs}\;, \label{detect}
\end{equation}
where $\chi_{obs}$ is the redshift-distance evaluated at the
maximum redshift ($z=z_{obs}$) of the survey used. We assume
throughout a $\Lambda$CDM model. In globally homogeneous
manifolds, $r_{inj}(x)$ is by definition position-independent, and
in this case it suffices to use the global injectivity radius
$r_{inj}$ (defined in general as $r_{inj}=\inf_{x\in
M}r_{inj}(x)$, which is the radius of the smallest sphere
inscribable in $M$) to determine sufficient conditions for
detectability. There are, however, significant classes of
$3-$manifolds which are not globally homogeneous, including the
totality of hyperbolic manifolds and the majority of the
multiply-connected spherical manifolds. One must therefore allow
for the fact that the detectability of cosmic topology may be
dependent on the observer's position.

Generally, a full systematic search for multiple images or pattern
repetitions of topological origin is a daunting task, as the very
diverse set of potential holonomies generate very different
patterns often dependent on the observer's position. Given,
however, that the fraction of the universe which is effectively
accessible to observations is limited, we shall in this work
concentrate only on detectable holonomies, that is those for
which at least two images of some radiating sources may be
observable. In this connection, we have shown in
Ref.~\cite{Mota-etal} that, assuming \noindent
\begin{itemize}
\item[{\bf (i)}]
The spatial sections of the Universe are not exactly flat;
\item[{\bf (ii)}]
The Universe has undergone a phase of inflationary expansion, such that
$|\Omega_{0}-1|~\ll 1,$ which ensures that $\chi_{obs}\ll 1$;%
\footnote{Here and in the following we express distances in units
of the curvature radius $a_0= |k|\, H_0^{-1} |\Omega_{0}-1|^{-1/2}$.}
\item[{\bf (iii)}]
The cosmic topology, or more properly some element of its holonomy
group, is detectable, i.e.,  Eq.~(\ref{detect}) holds for some point
$x \in M$;
\end{itemize}
then any detectable holonomy is (nearly) indistinguishable from a
Clifford translation (or CT, defined as a holonomy $\gamma$ under
which the distance between each point $x$ and its image $\gamma x$
is constant for all points in the manifold for most observers. As
a consequence, a generic compact non-flat manifold is ``locally''
well approximated by either a cylindrical ($\mathbb{R}^{2} \times
\mathbb{S}^{1}$) or toroidal ($\mathbb{R} \times \mathbb{T}^{2}$)
manifold, irrespective of its global topology.

In typical inflationary scenarios, achieving a sufficient number
of e-foldings required to generically solve the flatness and
horizon problems imposes bounds on the total density parameter
which is estimated as \cite{Linde2,TegmarkInf}
\begin{equation}
|\Omega_{0}-1|\ \lesssim\ 10^{-5}\ll 1 \;. \label{InfLimit}
\end{equation}
In what follows, we shall refer to this bound as the inflationary
limit.

For the bound given by Eq.~(\ref{InfLimit}) one can numerically calculate
$\chi_{obs}\,$. By using this value along with
Eq.~(\ref{detect}) we obtain an estimate of the detectability
condition in the inflationary limit
\begin{equation}
r_{inj}(x)\lesssim\chi_{obs} \lesssim 0.01 \;, \label{InfLim3}
\end{equation}
where we have taken $\Omega_{m0}=0.28$ and $z_{obs}= 1\,089$~\cite{wmap}.

\section{Bounds on the holonomy parameters}
\label{Sec3}

As was discussed in the previous section, an important
observational signature of any non-trivial detectable topology is
the existence of the circles-in-the-sky in the CMB maps, in which
matching pairs of circles are identified by a holonomy $\gamma \in
\Gamma$. Reciprocally, each detectable holonomy corresponds to a
pair of matching circles. Thus the detection of the matching
circles in the CMB maps would effectively amount to the detection
of the holonomy group (or some subgroup thereof), and in turn the
topology of the Universe. In general, for a given radius
$\chi_{obs}$ of the LSS, the relative position of any  pair of
matching circles depends on the specific holonomy which connects
the circles, the position of the observer and the value of
$\chi_{obs}$. In the case of Clifford translations, however, the
pairs of circles generated are always antipodal (or back-to-back)
for all observers and radii. A search that could be restricted to
such back-to-back pairs would of course be far less
computationally intensive than a full search covering the
entire parameter space.%
\footnote{A general search requires the computation of a
correlation function spanning a six-dimensional parameter space,
whose parameters give the location of the first and the second
circles centers [($\Theta_1,\Phi_1)$ and ($\Theta_2,\Phi_2)$,
say], their angular radius $\nu$ and the relative phase of the two
circles $\phi$.}
More generally, if one were able to somehow restrict the
permissible holonomy parameters from theoretical and observational
considerations, one could in turn restrict the parameter
space that needs to be spanned, thereby greatly simplifying the
search. This is what we aim to do here for most observers by
assuming the inflationary limit.

To obtain the holonomy parameters, we recall that the main outcome
of our earlier results~\cite{Mota-etal} is that, for most
observers (in the sense made precise below), the detectable
holonomies of nearly flat manifolds will deviate only by a small
amount from being Clifford translations (i.e., are CT-like), in
the sense that within the observable Universe the lengths
$d(x,\gamma x)$ of the closed geodesics generated by any
detectable holonomy $\gamma$ will be such that

\begin{equation}
\frac{\Delta d}{d_{0}}\leq\,\frac{2\chi_{obs}}{|z_{2}||z_{1}|} \; ,
\label{esfbound}
\end{equation}
and for hyperbolic manifolds
\begin{equation}
\frac{\Delta d}{d_{0}}\leq2\,\,\chi_{obs} \;. \label{bound3}
\end{equation}

In these expressions $\Delta d\equiv d_{\max}-d_{\min}$, where
$d_{\max}$ and $d_{\min}$ are respectively the maximum and minimum
lengths of the geodesics generated by this holonomy inside the
detectable sphere of radius $\chi_{obs}$, $d_{0}$ is the distance
function of the holonomy $\gamma$ evaluated at the center of the
sphere, and $z_{1}$ and $z_{2}$ are a pair of complex numbers
that parametrize the $3-$sphere (thus $|z_{1}|^{2}+|z_{2}|^{2}=1$),
which are used to express the holonomy $\gamma$ in its canonical form.
Of course, for true Clifford translations, $\Delta d /d_0=0$ (for more
details see Ref~\cite{Mota-etal}).

For all observers able to detect a holonomy $\gamma$ in hyperbolic
cusp-like manifolds, and for most such observers in a spherical
manifold (i.e., those bounded away from the equators, so that
$|z_{1}|,\ |z_{2}| \gg \chi_{obs}$), the bounds~(\ref{esfbound})
and (\ref{bound3}) imply that $\Delta d / d_0$ is very small,
which means that the holonomy is CT-like. It is easy to show that
the fraction of the volume of the manifold where $|z_{1}|<Z$ or
$|z_{2}|<Z$, for any $Z\leq1$, is proportional to $Z^{2}$. This
can be seen by writing $Z=\sin\xi_{max}$, $|z_{1}|=\cos\xi$, and
$|z_{2}|=\sin\xi$, where $0\leq\xi\leq\pi/2$ and
$0\leq\xi_{max}\leq\pi/4$. The holonomy group $\Gamma$ of manifold
$\mathbb{S}^{3}/\Gamma$ tiles the $3$-sphere $\mathbb{S}^{3}$ into
$\mathcal{O}(\Gamma)$  identical copies of the fundamental domain,
where $\mathcal{O}(\Gamma)$ is the order of $\Gamma$. Therefore,
the volume of the region where either $\xi\leq\xi_{\max}$ or,
conversely, $\xi\geq \pi/2-\xi_{\max}$, corresponds to the sum of
the volumes of two tori in the $3-$sphere
$|z_{1}|^{2}+|z_{2}|^{2}=1$ centered at $|z_{1}|=0$ and
$|z_{2}|=0$ (with radii $\xi_{\max}$ and $\pi/2-\xi_{\max}$
respectively), divided by $\mathcal{O}(\Gamma)$. In toroidal, or
Hopf, coordinates, this volume is given by
\begin{eqnarray}
V_{\xi_{\max}}\!\
&=&\frac{1}{\mathcal{O}(\Gamma)}\int_{0}^{2\pi}\!\!\!\int_{0}^{2\pi}\!\!\!
\left(\int_{0}^{\xi_{\max}}\!\! +
\int_{\pi/2-\xi_{\max}}^{\pi/2}\right)\! \cos\xi \nonumber \\
&\times& \,\sin\xi \, d\phi_{1}\,d\phi_{2}\,d\xi
               =\,\frac{4\pi^{2}Z^{2}}{\mathcal{O}(\Gamma)} \;.
\end{eqnarray}
Now, the volume of the manifold $\mathbb{S}^{3}/\Gamma$ is
given by the volume of the $3$-sphere divided by
$\mathcal{O}(\Gamma)$, i.e.,
$V_{{\mathbb{S}^{3}/\Gamma}}=V_{S^3}/\mathcal{O}(\Gamma)
=V_{\pi/2}=2\pi^{2} / \mathcal{O}(\Gamma)$.
Thus the fraction $\mathcal{R}$ of $\mathbb{S}^{3}/\Gamma$ where
$|z_{1}|\leq Z$ or $|z_{2}|\leq Z$ is given by
\begin{equation}
{\cal{R}} =V_{\xi_{\max}}/V_{\mathbb{S}^{3}/\Gamma}=2Z^{2} \;.\label{VV}
\end{equation}
It is then clear from (\ref{VV}) that the volume of the region
where $z_{1}\sim\chi_{obs}$ or $z_{2}\sim\chi_{obs}$, as a
fraction of the volume of the manifold, is of the order
of $2\chi_{obs}^{2}$, which is very small if the inequality
$\chi_{obs} \ll 1$ holds.
The smallness of $\cal{R}$ then implies that
the probability of finding an observer
in such regions is small (in the same sense as that
defined by the so called injectivity profiles~\cite{Weeks}),
which means that~(\ref{esfbound}) imposes stringent bounds
on $\Delta d / d_0$ in all but a small region of the $3$-sphere
whose volume is of the order of $\chi_{obs}^{2}$.
In what follows, therefore, we shall confine
ourselves to the case of \emph{typical observers},
i.e. those that reside away from these small probability
regions.

Now, in the limit~(\ref{esfbound}) and (\ref{bound3}), the
geometry of the (non-flat) spatial sections of the observable
universe will be well-approximated by the geometry of flat space
$\mathbb{R}^3$. Therefore, any detectable holonomy will likewise
be locally well-approximated by some isometry of $\mathbb{R}^3$.
It is known, however, that any orientation-preserving flat
isometry (and, in particular, any flat holonomy) can be expressed
as a screw motion~\cite{Thurston}, which consists of a rotation
around a suitable axis followed by a translation along the same
axis. In what follows, we shall write a generic non-flat
detectable holonomy $\gamma$ in the limit~(\ref{esfbound}) and
(\ref{bound3}) as a generic screw motion, and use our previous
bounds on nearly flat detectable holonomies to constrain the
parameters characterizing such holonomy.

\begin{figure}[tb!]
\begin{center}
\includegraphics[width=8.2cm,height=5.2cm,angle=0]{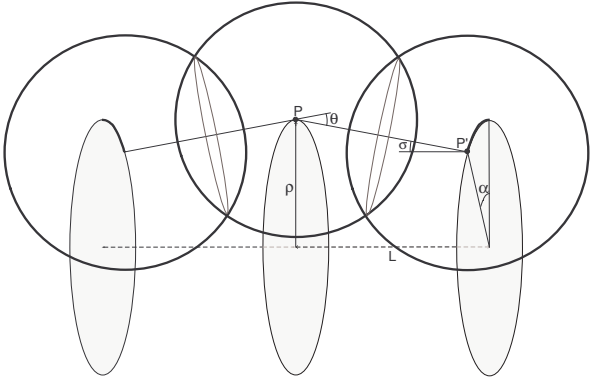}
\caption{\label{Fig1} Depiction of a screw motion isometry
$\gamma$, whose action takes point the $P$ to $P^{\prime}$. This
amounts to a translation by $L$ along the axis of isometry (dotted
line), and a rotation by $\alpha$ around the same axis. The LSS
sphere (thick solid line) centered at $P$ intersects its images
(also in thick solid lines) centered at $\gamma P$ and
$\gamma^{-1}P$ along two matched circles-in-the-sky. }
\end{center}
\end{figure}

Let us consider, without loss of generality, a typical observer at
a point $P$ (see Fig. 1) with coordinates chosen such that
$P=(\rho,0,0)$, where $\rho$ is the distance from the axis of
rotation. Let the image of the point $P$ under the action of a
holonomy $\gamma$ be $P^{\prime}=\gamma P$ with coordinates
$(\rho\cos\alpha,\rho\sin\alpha,L)$, where $\alpha$ is the phase
angle and $L$ is the translation length corresponding to the screw
motion isometry. The length of the
closed geodesic connecting $P$ and $P^{\prime}$ is given by
\begin{align}
d_{0}  &  =\left\vert P^{\prime}-P\right\vert =\left\vert \rho(1-\cos
\alpha),\ \rho\sin\alpha,\ L\right\vert \nonumber\\
&  =\sqrt{2\rho^{2}\left(  1-\cos\alpha\right)  +L^{2}} \;.\label{length}%
\end{align}

To proceed, we need to calculate not only the length of the
geodesic passing through the observer's position given by
Eq.~(\ref{length}), but also the lengths of the longest and
shortest closed geodesics within the observable Universe,
which are respectively given by
\begin{align}
d_{\max}  &  =\sqrt{2\left(  \rho+\chi_{obs}\right)^{2}(1-\cos\alpha)+L^{2}%
} \label{length2a}\;, \\
\nonumber \\
d_{\min}  &  =\!\!\left\{
\begin{array}
[c]{l}%
\!\!\! \sqrt{2\left(  \rho-\chi_{obs}\right)^{2}(1-\cos\alpha)+L^{2}}%
\;\;\mbox{if} \;\ \ \rho\geq\chi_{obs} \,,\\
L \quad \quad \quad \qquad \qquad \qquad \qquad \qquad \mbox{if} \;\;\; \rho\leq
\chi_{obs} \label{length2b}\,.
\end{array}
\right.
\end{align}

Since $L\leq [2(\rho-\chi_{obs})^{2}(1-\cos\alpha)+L^{2}]^{1/2}$~%
holds~identically,~it~follows~that $d_{\min} \leq [2(\rho
-\chi_{obs})^{2}(1-\cos\alpha)+L^{2}]^{1/2}$. Thus, combining
Eqs.~(\ref{length})~--~(\ref{length2b})
we obtain
\begin{widetext}
\begin{equation}
\frac{\Delta
d}{d_{0}}\geq\sqrt{1+\frac{1}{d_{0}^{2}}(4\rho\chi_{obs}+2\chi
_{obs}^{2})(1-\cos\alpha)} \, - \,
\sqrt{1+\frac{1}{d_{0}^{2}}(-4\rho\chi_{obs}
+2\chi_{obs}^{2})(1-\cos\alpha)}\;.
\end{equation}
\end{widetext}
The terms on the right hand side are respectively greater and
smaller than $1$. Also, according to (\ref{esfbound}) and
(\ref{bound3}), the difference between the two terms must be $ \ll
1$. Therefore, both terms must be $\sim 1$, which allows them to
be expanded to obtain
\begin{equation}
\frac{\Delta
d}{d_{0}}\geq\frac{4\rho\chi_{obs}}{d_{0}^{2}}(1-\cos\alpha
)+\mathcal{O}\,(\alpha^{4})\;. \label{9}
\end{equation}
Note that since $4\,\rho\, \chi_{obs} / d_{0}^{2} \gg 1$, then in order to keep
the first term $<1$ we must have $(1-\cos\alpha)\ll1$, which implies that
$\alpha$ is very small.

Now let the angle between the axis of the screw motion and the
segment $(P^{\prime}-P)$ be $\sigma$ (see Fig. \ref{Fig1}), such
that $\cos\sigma= L/d_{0}$. It can then be shown that
\begin{equation}
\tan\sigma=\frac{\rho}{L}\sqrt{2(1-\cos\alpha)} \;. \label{tansigma}
\end{equation}
Using (\ref{tansigma}) to express $\alpha$ in terms of $\sigma$,
it is possible to write equation (\ref{9}) as
\begin{equation}
\frac{\Delta
d}{d_{0}}\geq\frac{2\chi_{obs}}{\rho}\sin^{2}\sigma+\mathcal{O}\,
(\alpha^{4})\;.
\end{equation}

Combining the above results with the bounds~(\ref{esfbound}) and
(\ref{bound3}) we finally obtain
\begin{align}
\,\,\frac{\rho}{|z_{2}||z_{1}|}\,  &  \geq\sin^{2}\sigma\;,
\quad \mbox{for spherical manifolds} \label{bound2a} \;,\\
\,\,\rho\,  &  \geq\sin^{2}\sigma\;,
\quad \mbox{for hyperbolic manifolds} \label{bound2b} \;.
\end{align}

\section{Bounds on the parameters of the circles-in-the-sky}
\label{Sec4}

\begin{figure}[tb!]
\begin{center}
\includegraphics[width=5.7cm,height=5.7cm,angle=0]{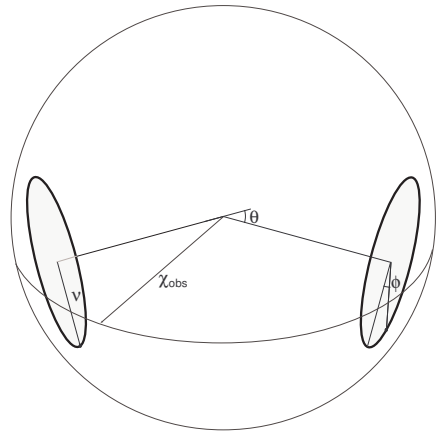}
\caption{\label{Fig2} This figure depicts the circles at the intersection
of the LSS with its images. The parameter $\theta$ is a measure of the
deviation of the circles from being antipodal, and $\phi$ measures the
phase difference between the circles of radius $\nu$.}
\end{center}
\end{figure}

In the previous section we used our previous bounds (c.f.~\cite{Mota-etal})
on $\Delta d / d_0$ to constrain the screw motion parameters
$(\rho,\alpha,L)$. To relate these to the corresponding parameters
$(\theta,\nu,\phi)$ for the circles-in-the-sky (see Fig.~\ref{Fig2}),
one needs the relations between the circles-in-the-sky and the
screw motion parameters, which are given by~\cite{German}

\begin{equation} 
\cos\theta=1-\sin^{2}\sigma(1-\cos\alpha) \;, \label{costheta}
\end{equation}
and \vspace{-0.4cm}
\begin{equation}
\cos\nu=\frac{L}{2\,\chi_{obs}\cos\sigma} \, \;. \label{cosni}
\end{equation}

Now, in the inflationary limit $\alpha$ is small, which according
to (\ref{costheta}) implies that $\theta$ is small. We can then
write, up to the second order in $\theta$,
\[
\theta=\frac{1}{\sqrt{2}}\sin\sigma\sqrt{1-\cos\alpha} \;.
\]
Using (\ref{tansigma}) and employing the bounds on $\sin\sigma$ from
(\ref{bound2a}) and (\ref{bound2b}), together with the expression~(\ref{cosni})
for $\cos\nu$, we obtain
\begin{align}
\,\,\theta &  \leq\frac{\sqrt{2} \cos\nu}{|z_{2}||z_{1}|}\,\chi_{obs}
\quad \mbox{for spherical manifolds} \label{citsA}\;, \\
\,\,\theta &  \leq\sqrt{2} \cos\nu \,\chi_{obs}
\quad \mbox{for hyperbolic manifolds} \label{citsB} \;.
\end{align}

Clearly, these inequalities provide upper bounds on the values of
the angle $\theta$ that characterizes the deviation from
\emph{antipodicity} of pairs of circles-in-the-sky as a function
of the circles' radii $\nu$, the distance to the LSS $\chi_{obs}$,
and the observer's position (for the spherical case). The
dependence of these bounds on $\theta$ on the density parameters can
be made explicit  by recalling that for the
small values of $|\Omega_{0}-1|$ given by the bound~(\ref{InfLimit}),
the contour curve $\chi_{obs}(\Omega_{m0},\Omega_{\Lambda0})=r_{inj}$
can be well-approximated by the secant line joining its intersections
with the $\Omega_{m0}=0$ and $\Omega_{\Lambda0}=0$
axes~\cite{Mota_etal2003}. Using this approximation we
obtain
\begin{equation}
\chi_{obs}\simeq 2\,\sqrt{\frac{|\Omega_{0}-1|}{\Omega_{m0}}} \;,
\label{analit35}
\end{equation}
which can be substituted in inequalities~(\ref{citsA}) and
(\ref{citsB}) to give
the bounds
\begin{equation}
\theta   \lesssim \,\frac{2\sqrt{2} \cos\nu}{|z_{2}||z_{1}|} \,\,
\sqrt{\frac{|\Omega_{0}-1|}{\Omega_{m0}}} \;,
\end{equation}
and
\begin{equation}
\theta   \lesssim  \,2 \sqrt{2} \cos\nu \,\,
\sqrt{\frac{|\Omega_{0}-1|}{\Omega_{m0}}} \;,
\end{equation}
for, respectively, spherical and hyperbolic manifolds.

Thus given observational bounds on the density parameters they
allow constraints to be set on the parameter $\theta$ for any
radii $\nu$ of matching circles to be considered in order to
perform a comprehensive search for matching circles in the
observable Universe.

Finally, as is clear from our discussion in Section~\ref{Sec3},
there are some observers in spherical manifolds --- namely those
close to the equators ($z_1 \simeq 0$ or $z_2\simeq 0$) --- for
whom the bounds on $\theta$ derived here are not applicable.
However, as can be seen from Eq. \ref{VV}, the set of such
observers is very small in the inflationary limit.

\section{Implications of our results and searches for circles-in-the-sky}
\label{Sec5}

To illustrate the application of the bounds~(\ref{citsA}) and%
~(\ref{citsB}), we consider the recent search for
circles-in-the-sky using the WMAP data~\cite{CitSRedux}, in order
to determine the sets of topologies that can be ruled out. Given
the prohibitive numerical cost of a full search, this search was
confined to antipodal or nearly-antipodal circles with deviation
from antipodicity $\theta \le 10^{\circ}$ and with radii
$\nu\ge18^{\circ}$. This search found no statistically significant
pairs of matching circles.

\begin{figure}[b!]
\begin{center}
\includegraphics[width=7.8cm,height=7.2cm,angle=0]{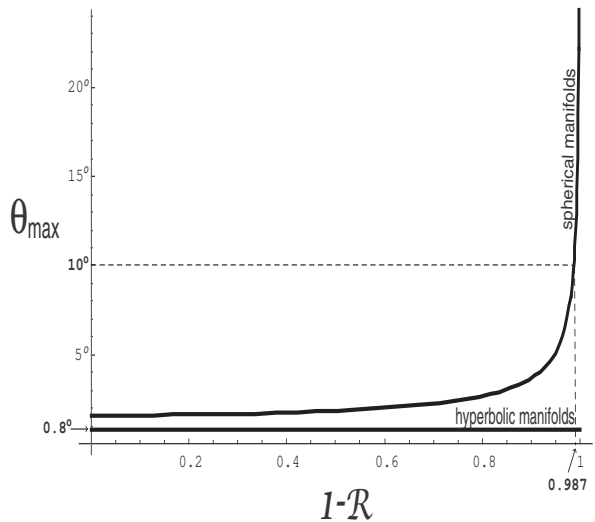}
\caption{\label{Fig3} Upper bound $\theta_{\max}$ for the maximum
deviation from antipodicity as a function of the fraction of
potential observers for which $\theta \leq \theta_{\max}$ in the
inflationary limit. This figure shows, for example, that for 98.7\%
of all observers in spherical manifolds, $\theta \leq 10^{\circ}$.
It also shows that for all observers in hyperbolic manifolds,
$\theta \leq 0.8^{\circ}$. In this figure, we have taken
$\Omega_{m0}=0.28$ and circles with radii $\nu \geq 18^{\circ}$,
but for circles with arbitrarily small radii, the value of
$\theta_{\max}$ would increase by a factor of at most $1.05$.}
\end{center}
\end{figure}

\begin{figure}[bth!] 
\begin{center}
\includegraphics[width=8cm,height=7.4cm,angle=0]{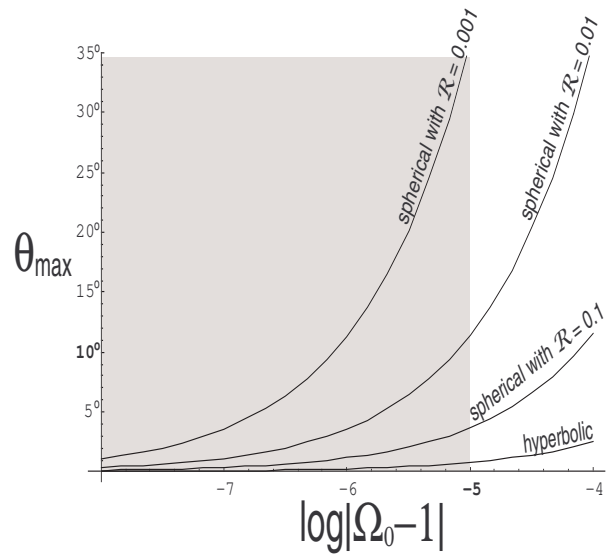}
\caption{\label{Fig4}Upper bound $\theta_{\max}$ for the maximum
deviation from antipodicity as a function of the total density
parameter $\Omega_0$, calculated for the sets of 90\%, 99\% and
99.9\% of observers (respectively, ${\mathcal R}=0.1,0.01$ and
$0.001$), and for the totality of observers in hyperbolic
manifolds. The shaded region corresponds to the inflationary limit
as defined in the text. Again, we have taken $\Omega_{m0}=0.28$
and circles with radii $\nu \geq 18^{\circ}$.}
\end{center}
\end{figure}

If confirmed, an important question regarding this search would be to
determine concretely whether it is sufficient to rule out
detectable non-trivial, non-flat, cosmic topologies, assuming the
inflationary limit~(\ref{InfLimit}). To answer this question we
have, employing the inequalities~(\ref{citsA}) and~(\ref{citsB}),
plotted in Fig.~\ref{Fig3} an upper bound, $\theta_{\max} $, for
the maximum deviation from antipodicity of a pair of matching
circles as a function of the fraction of the observers in
spherical and hyperbolic manifolds. As can be seen from this
figure, and concretely estimated from inequalities~(\ref{citsA})
and~(\ref{citsB}), in the inflationary limit~(\ref{InfLimit}), the
angle $\theta_{\max}$ is less than $0.8^{\circ}$ for any observer
in a hyperbolic manifold.%
\footnote{We note, however, that for most such observers
the topology is undetectable to begin with (see Ref.~\cite{Weeks}
for details).}
On the other hand, for $98.7\%$ of the observers in spherical
manifolds $\theta_{\max} \le 10^{\circ}$. In both cases we have
confined ourselves to circles with radii $\nu \ge 18^{\circ}$
(following~\cite{CitSRedux}); but had we considered circles of
arbitrarily small radii, the upper bounds on $\theta$ would
increase only by a factor of $1.05$. This implies that, if one
accepts the negative result of the search for circles of
Ref.~\cite{CitSRedux}, then the assumption of inflationary limit
considered here is sufficient to rule out detectable non-trivial
cosmic topologies for overwhelming majority of the potential
observers. Thus to detect a non-trivial topology an observer must
either live in a restricted region of some spherical manifolds or
the Universe must have a total density parameter which is not too
close to the critical density. This latter statement can be made
precise by plotting $\theta_{\max}$ as a function of the density
parameter ($|\Omega_0 -1|$) for different sets of observers
(Fig.~\ref{Fig4}). So, for instance, the search undertaken in
Ref.~\cite{CitSRedux} is sufficient to rule out a detectable
cosmic topology for all observers in hyperbolic manifolds, and for
at least $99.9\%$ of the observers in spherical manifolds, if
$|\Omega_0 -1| \le 10^{-6}$. On the other hand, if $|\Omega_0 -1|
\le 10^{-4}$, then such a detectable topology is still
definitively excluded for all observers in hyperbolic manifolds,
but only for $90\%$ of observers in spherical manifolds.

Future searches for pairs of correlated circles may of course
study different regions of the circles-in-the-sky parameter space.
As Fig.~\ref{Fig4} illustrates, for all observers in hyperbolic
manifolds, even searches restricted to small values of $\theta$
(if confirmed) would be sufficient to exclude a detectable non-trivial
cosmic topology for relatively large values of total density, i.e.
for $1-\Omega_{0}\sim10^{-4}$.
On the other hand, the fraction of
observers in spherical manifolds that may observe pairs of circles
with $\theta \geq \theta_{max}$ varies significantly with the
value of $\theta_{max}$. Thus, for example, a search for pairs
with $\theta \leq 5^{\circ}$ would be able to exclude a detectable
non-trivial cosmic topology for $99.9 \%$ of observers only if
$\Omega_0 -1 \le 1.2 \times 10^{-7}$, a value almost one order of
magnitude smaller than that for $\theta \leq 10^{\circ}$.
Conversely, it would take a search for $\theta\leq 35^{\circ}$ to
exclude a detectable non-trivial topology for the same $99.9 \%$
of observers  for the range of values of $\Omega_0$ given
by the inflationary limit (\ref{InfLimit}).

Of course, for sufficiently low values of $|\Omega_{0}-1|$, a
detectable non-trivial (non-flat) topology can be excluded for an
arbitrarily small fraction of potential observers and arbitrarily
small values of $\theta_{max}$. However, one can not completely
exclude all potentially detectable spherical holonomies for all
observers, since one can always find one such holonomy that is
well-approximated in some neighborhood around the equator $z_1=0$
by a screw motion of arbitrarily large twist, which will in turn
generate, for some values of $\chi_{obs}$, circle pairs with
arbitrary deviation from antipodicity.

A non-detection in a limited search for pairs of correlated
circles would likewise not rule out the possibility of a
detectable flat non-trivial topology. Since flat holonomies lack a
characteristic scale factor, there is some considerable leeway in
choosing the holonomy parameters, and it is easy to obtain
holonomies that produce circle pairs of arbitrary $\theta$ for
typical observers. Indeed, if one assumes the inflationary limit,
then the detection of a pair of correlated circles with high
$\theta$ would strongly suggest that the Universe has flat spatial
sections, since pairs of correlated circles with high values of
$\theta$ would be present for no observers in a hyperbolic
Universe, and for only a small fraction of observers in a
spherical Universe.

Although it is not possible to exclude the possibility of
detection of the cosmic topology by all observers, the combination
of such negative results of any search (present or forthcoming),
together with the results derived in this paper, would allow
precise bounds to be put on the fraction of the potential
observers for whom a non-trivial cosmic topology would be ruled
out, for any given value of the density parameters.

\section{Final Remarks}
\label{Sec6}

An important model-independent observational signature of a
detectable non-trivial cosmic topology is the occurrence of pairs
of matching circles of temperature fluctuations in maps of the
cosmic microwave background radiation. Here, by employing some recent
results concerning the local nature of generic non-flat detectable
non-trivial topology in the inflationary limit, we have obtained
concrete bounds on the angular parameter characterizing the deviation
from antipodicity of circles-in-the-sky as a function of the cosmological
density parameters and the position of the observer, for any radius
$\nu$ of the circles.

As an example of the application of our results, we have
considered the most recent search restricted to nearly
back-to-back circles, which has found no statistically significant
pairs of matching circles. Using our bounds we have found that,
assuming the total density parameter satisfies $0<|\Omega_{0}-1|
\lesssim10^{-5}$, this search, if confirmed, could in principle be
sufficient to exclude a detectable non-trivial cosmic topology for
most observers.

Our results also provide a framework to draw quantitative
conclusions from the negative results of such partial searches,
past and future, for circles-in-the-sky as a generic signature of
non-trivial cosmic topology. More specifically, they allow us to
quantify the fraction of the potential observers for which the
absence  of pairs of matching circles-in-the-sky in CMB maps rules
out a non-trivial (non-flat) topology for the spatial sections of
the Universe, as a function of the  cosmological density
parameters. We emphasize that these results apply generically to
all potential non-flat manifolds with non-trivial topology
rather than specific classes. In particular, if the negative
results of the recent searches are confirmed,  then in
the inflationary limit a detectable non-trivial (non-flat)
topology is excluded for all observers apart from a very
small subset, if the Universe has positively-curved spatial
sections, and for all observers if the spatial sections
turn out to be negatively-curved.
%
\begin{acknowledgments}
This work is supported by Conselho Nacional de Desenvolvimento
Cient\'{i}fico e Tecnol\'ogico (CNPq) -- Brasil, under grant No.\
472436/2007-4. M.J. Rebou\c{c}as and R. Tavakol thank CNPq and
PCI-CBPF/MCT/CNPq for the grants under which this work was carried
out. We would also like to thank G. Gomero for fruitful discussions.
\end{acknowledgments}


\end{document}